\documentclass[journal]{IEEEtran}
\usepackage{amsmath}
\usepackage{graphicx}
\usepackage{amsfonts}
\usepackage{amsmath,epsfig}
\usepackage{stfloats}
\usepackage{amssymb}
\usepackage{cite}
\ifCLASSINFOpdf

\else

\fi

\begin{document}
%
\title{Efficient Linear Transmission Strategy for MIMO Relaying Broadcast Channels  with Direct Links }

\author{Haibin~Wan, Wen~Chen,~\IEEEmembership{Senior~Member,~IEEE,}
        and Jianbo~Ji
\thanks{Manuscript received October 7, 2011; accepted November 17th, 2011. The associate editor coordinating the review of this paper
and approving it for publication was H. Viswanathan.}
\thanks{H.~Wan, W.~Chen, and J.~Ji are with Department of Electronic Engineering, Shanghai Jiao Tong University,
China; H.~Wan is also with School of Physics Science and Technology,
Guangxi University,
China; W.~Chen is also with the SKL for ISN, Xidian University,
China~(e-mail:\{dahai\_good;wenchen;jijianbo\}@sjtu.edu.cn)}

\thanks{This work is supported by NSF China \#60972031, by national 973 project \#2012CB316106 and
\#2009CB824904, by national huge special project \#2012ZX03004004,
by national key laboratory project \#ISN11-01, by Huawei Funding
\#YBWL2010KJ013, and by Foundation of GuangXi
University~\#XGL090033.}
 }
\markboth{IEEE Wireless Communications Letters ,~Vol.~1, No.~1, January~2012}%
{Shell \MakeLowercase{\textit{et al.}}: Bare Demo of IEEEtran.cls for Journals}

\maketitle

\begin{abstract}
In this letter, a novel linear transmission strategy to design the
linear precoding  matrix~(PM)  at base station~(BS) and the
beamforming matrix~(BM) at relay station~(RS) for multiple-input
multiple-output~(MIMO) relaying broadcast channels with direct
channel (DC) is proposed, in which a linear PM is designed  at BS
based on DC, and the RS utilizes the PM, the backward channel and the forward channel  to design the linear BM. We then give a
quite tight lower bound of the achievable sum-rate of the network
with the proposed strategy to measure the performance. The sum-rates
achieved by the proposed strategy is compared with other schemes
without considering the DC in design in simulations, which shows
that the proposed strategy outperforms the existing methods when RS
is close to BS.
\end{abstract}
\IEEEpeerreviewmaketitle

\section{Introduction}
\IEEEPARstart{R}{ecently}, MIMO relaying broadcast network has attracted considerable interest from both academic and industrial communities. For a
MIMO relaying broadcast network, there are two independent channels
between source and destination nodes; i.e., relay channel consisting
of \emph{backward channel} (BC) and \emph{forward channel} (FC), and
\emph{direct channel} (DC).  Many
works~\cite{2008-Chae-fixedRelay,2009-Rui-Zhang-Qos,2010-WeiXu,2011-Wei-Xu,2010-Gomadam,2010-BinZhang,2010-WeiXu-Feedback}
 have investigated the linear transmission strategy for MIMO relaying broadcast networks.
In~\cite{2008-Chae-fixedRelay}, an implementable multiuser precoding strategy
that combines Tomlinson-Harashima precoding at the BS and linear signal processing at the RS is presented.
In~\cite{2009-Rui-Zhang-Qos}, a joint optimization of linear
beamforming and power control at BS and RS  to minimize  the
weighted sum-power consumption under the user minimum
SINR-(QoS)-constraints is presented.
In~\cite{2010-WeiXu}, the singular value decomposition (SVD) and
zero  forcing~(ZF) precoder are respectively used to the BC and FC
to optimize the joint precoding. The authors use an iterative method
to show that the optimal precoding matrices always diagonalize the
compound channel of the system.
In~\cite{2011-Wei-Xu}, the authors use the quadratic programming to
joint precoding optimization to maximize the system capacity.
In~\cite{2010-Gomadam}, the authors   propose a scheme based on
duality of MIMO multiple access and broadcast channel to maximize
the system capacity.
In~\cite{2010-BinZhang} and~\cite{2010-WeiXu-Feedback},  the authors
consider  a robust linear beamforming scheme with a limited
feedback.

However, all these works only consider the BC and FC to  optimize
precoding matrix (PM) and beamforming matrix (BM) to maximize the
system performance. For a MIMO relay  network with DC, jointly
designing the PM and  BM to maximize capacity or minimize the mean
square error  is  much difficult, especially for a MIMO relaying
broadcast  networks.
In this letter, we  consider a MIMO relaying  broadcast networks
composing of one  BS, one RS and multiple users with DC, and propose
a novel linear transmission strategy to design the PM  and BM. To
avoid  complexity, we  consider the distributed design method.
Firstly, the BS  designes the  linear PM based on DC only, Secondly,
the RS utilizes the linear PM, BC and FC to design the BM.
Simulation results demonstrates that the proposed strategy
outperforms the existing methods when  RS is close to BS.
\begin{figure}[!t]
\begin{center}
\includegraphics [width=3.2in]{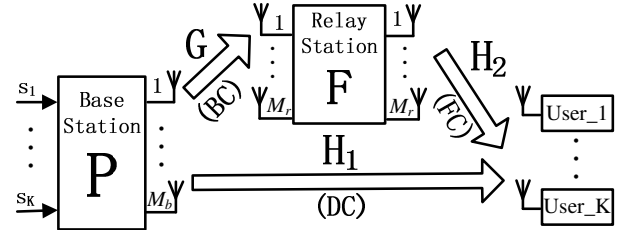}
\caption{The MIMO relaying broadcast network  with one base station,
one relay station, and $K$ mobile users} \label{Fig-Model}
\end{center}
\end{figure}

\emph{Notations:} $\textmd{E}(\cdot)$, $\mathrm{tr}(\cdot)$, $(\cdot)^{T}$, $(\cdot)^{-1}$, $(\cdot)^{\dag}$, and $\mathrm{det}(\cdot)$ denote expectation, trace, inverse, transpose,  conjugate transpose, and determinant, respectively. \emph{i.i.d } stands for independent and identically distributed. $\mathbf{I}_N$ stands for  an $N\times N$ identity matrix. $\mathrm{diag}(a_{1},\ldots,a_{N})$ is a diagonal matrix with the $i$th diagonal entry $a_{i}$. $\log$ is of base $2$. $\mathcal{C}^{M\times N}$ represents the set of $M\times N$ matrices over complex field, and $\sim\mathcal{CN}(x,y)$ means satisfying  a circularly symmetric complex Gaussian distribution with  mean $x$ and covariance $y$.
\section{System Model}
In this letter, we consider the MIMO relaying   broadcast channel
with  one BS, one RS, and $K$ single-antenna users as depicted in
Fig~\ref{Fig-Model}. It is assumed that the BS and RS are equipped
with $M_b$ and $M_r$ antennas,  respectively. We consider that each
user can receive the signals from the BS via DC and BC-FC. Assuming
that the BS can support $K$ independent substreams for $K$ users
simultaneously,  which requires $K\leq M_b$. Here, we only consider
$M_b=M_r=K$ for simplicity. By means of relaying scheme, we consider
a non-regenerative and half-duplex relaying scheme applied at the RS
to process and forward the received signals due to
simplicity~\cite{2004-Laneman}. Thus, a broadcast transmission is
composed of two phases. During the first phase, the BS broadcasts
$K$ precoded data streams to the RS and users after applying a
linear PM $\mathbf{P}$ to the  data vector $\mathbf{s}$, where
$\mathbf{s}=[s_1,s_2,\ldots,s_K]\sim\mathcal{CN}(0,\mathbf{I}_{K})$,
$s_k$ is the symbol intended for the $k$th user. We supposed that
the BS transmit power is $P_s$ and then the  power control factor at
the BS is
$\rho_{s}=\sqrt{{P_s}/{\mathrm{tr}(\mathbf{P}\mathbf{P}^{\dag})}}$
due to
$\textmd{E}\{\mathbf{Ps}\mathbf{s}^{\dag}\mathbf{P}^{\dag}\}=\mathrm{tr}(\mathbf{P}\mathbf{P}^{\dag})$.
In this case, the received signal vectors at the users and RS are
respectively
\begin{IEEEeqnarray} {rCl} \label{eq:Downlink-BS}
\mathbf{y}_{1} &=&[y_{11},y_{12},\ldots,y_{1K}]^{T}=\rho_{s} \mathbf{H}_{1}\mathbf{P}\mathbf{s } +\mathbf{n}_{1}\IEEEyessubnumber \label{eq:Downlink-y1},\\
\mathbf{y}_{r} &=&\rho_{s} \mathbf{G}\mathbf{P}\mathbf{s } +\mathbf{n}_{r},\IEEEyessubnumber \label{eq:Downlink-yr}
\end{IEEEeqnarray}
where
$\mathbf{P}=[\mathbf{p}_{1},\mathbf{p}_{2},\ldots,\mathbf{p}_{k}]\in
\mathcal{C}^{M_b\times K}$ is a PM at BS, and
$\mathbf{H}^{T}_{1}=[\mathbf{h}_{11},\mathbf{h}_{12},\ldots,\mathbf{h}_{1k}]\in
\mathcal{C}^{K\times M_b}$ is the DC matrix, in which
$\mathbf{h}_{1k}$ is the channel vector between BS and the $k$th
user. $\mathbf{G}$ is the BC matrix. Each entry in $\mathbf{H}_{1}$
and $\mathbf{G}$ is \emph{i.i.d} complex Gaussian variables with
zero mean and variances $\sigma^2_{h_1}$ and  $\sigma^2_{g}$,
respectively. $\mathbf{n}_{i}=[n_{i1},n_{i2},\ldots,n_{ik}]^{T}$
$\sim\mathcal{CN}(0,\sigma^2_{n}\mathbf{I}_{K})$ ($i=1,r$) is the
noise vectors at users and RS. $y_{1k}$ and $\mathbf{y}_{r}$ are the
received signal symbol at the $k$th user and the received signal
vector at RS, respectively.

During the second phase,  the RS forwards the received signal vector
to the users after  a linear BM $\mathbf{F}$. The transmit power at
the RS is $P_r$, and the relay power control factor  $\rho_r$ is
\begin{equation}\label{eq:rho-r}
\rho_{r}=\sqrt{{P_r}/{\mathrm{tr}(\rho^2_{s}\mathbf{F}\mathbf{G}\mathbf{P}\mathbf{P}^{\dag}\mathbf{G}^{\dag}\mathbf{F}^{\dag} +\sigma^2_{n}\mathbf{FF}^{\dag})}}.
\end{equation}
Denoting the received signal symbol at the $k$th user as $y_{2k}$,
the received signal vector at users can thus be written as
\begin{equation} \label{eq:Relay-Downlink}
\mathbf{y}_{2} =[y_{21}~\cdots~ y_{2K}]^{T}=\rho_{r}\rho_{s} \mathbf{H}_{2} \mathbf{F }\mathbf{G}\mathbf{P}\mathbf{s } +\rho_{r}\mathbf{H}_{2}\mathbf{F}\mathbf{n}_{r}+\mathbf{n}_{2},
\end{equation}
where
$\mathbf{H}^{T}_{2}=[\mathbf{h}_{21},\mathbf{h}_{22},\ldots,\mathbf{h}_{2k}]\in
\mathcal{C}^{K\times M_r}$   is a MIMO BC matrix, in which
$\mathbf{h}_{2k}$ is the channel vector between RS and the $k$th
user. Each entry in $\mathbf{H}_{2}$ is \emph{i.i.d} complex
Gaussian variables with zero mean and variances $\sigma^2_{h_2}$.
$\mathbf{n}_{2}=[n_{21},n_{22},\ldots,n_{2k}]^{T}\sim\mathcal{CN}(0,\sigma^2_{n}\mathbf{I}_{K})$
is the noise vector at user.

Overall, we model the MIMO relaying broadcast network with DC as two
broadcast channels in (\ref{eq:Downlink-y1}) and
(\ref{eq:Relay-Downlink})  with power constraints at the BS and the
RS being $P_s$ and $P_r$, respectively.

\section{Base Station  Precoding and Relay Beamforming  Design}
In this section,   we choose a regularized ZF~(RZF)\cite{2005-RZF}
precoder at BS based on DC, and then propose a novel  linear
beamforming scheme based on ZF-RZF beamforming techniques at RS
within the context of  the proposed strategy.
\subsection{Base Station Precoding Design Based on RZF }
The design of the PM at the BS depends on the user operation mode.
For the BS-users direct mode and BS-RS-users relay mode, the
precoding design have been well investigated in several
papers~\cite{2003-Caire-DPC,2005-RZF,2007-Mirette,2010-Xiyuan-Wang}
and
\cite{2008-Chae-fixedRelay,2009-Rui-Zhang-Qos,2010-WeiXu,2010-Gomadam,2011-Wei-Xu,2010-BinZhang}.
However, to our best knowledge, there is no work investigating the
mixed mode in MIMO relaying broadcast network with DC. In this
letter, we consider the PM at BS as like the direct mode to make
better use of DC. We assume that the BS has perfect CSI of DC. We
choose an RZF filter as the PM for the BS due to it is linear and
can achieve near-capacity at sum-rate~\cite{2005-RZF}. Therefore,
the PM is
\begin{equation}\label{eq:matix-P}
\mathbf{P}=\mathbf{H}^{\dag}_{1}(\mathbf{H}_{1}\mathbf{H}^{\dag}_{1}+\alpha \mathbf{I})^{-1},
\end{equation}
where the optimal $\alpha$ is equal to the ratio of total noise variance to the total transmit power, i.e., $\alpha_{opt}=K\sigma^{2}_{n}/P_s$~\cite{2005-RZF}. Consequently,  the received  signal vectors (\ref{eq:Downlink-BS}) at the users and RS can be rewritten as
\begin{IEEEeqnarray} {rCl} \label{eq:Downlink-BSP}
\mathbf{y}_{1} &=& \rho_{s} \mathbf{H}_{1}\mathbf{H}^{\dag}_{1}(\mathbf{H}_{1}\mathbf{H}^{\dag}_{1}+\alpha \mathbf{I})^{-1}\mathbf{s } +\mathbf{n}_{1} \IEEEyessubnumber \label{eq:Downlink-P-y1},\\
 \mathbf{y}_{r} &=& \rho_{s} \mathbf{G}\mathbf{H}^{\dag}_{1}(\mathbf{H}_{1}\mathbf{H}^{\dag}_{1}+\alpha \mathbf{I})^{-1}~\mathbf{s } +\mathbf{n}_{r}. \IEEEyessubnumber \label{eq:Downlink-P-yr}
\end{IEEEeqnarray}
\subsection{Relay Beamforming  Design Based on ZF-RZF}
To design a BM at RS, we assumed that the RS has known the
$\mathbf{P}$ and the perfect knowledge  of  $\mathbf{G}$ and
$\mathbf{H}_2$. To design  an efficient BM at RS, we divide the BM
$\mathbf{F}$ into two parts: $1)$ the receiving BM $\mathbf{F}_R$,
and $2)$ the transmitting BM $\mathbf{F}_T$, i.e.,
$\mathbf{F}=\mathbf{F}_{R}\mathbf{F}_{T}$. The transmitting BM
$\mathbf{F}_T$ is like the PM at BS.  We also choose an RZF filter
as the transmitting BM for the RS. Thus, the transmitting BM
$\mathbf{F}_T$ can be written as
\begin{equation}\label{eq:F-t}
\mathbf{F}_{T}=[\mathbf{f}_{T1},\mathbf{f}_{T2}\ldots,\mathbf{f}_{TK}]=\mathbf{H}^{\dag}_{2}(\mathbf{H}_{2}\mathbf{H}^{\dag}_{2}+\gamma \mathbf{I})^{-1},
\end{equation}
where the optimal $\gamma$ is $\gamma_{opt}=K\sigma^{2}_{n}/P_r$~\cite{2005-RZF}.

The next work is to decide the receiving BM $\mathbf{F}_R$. In this
letter, we utilize a ZF filter~\cite{2005-DavidTse-book}  as the
receiving BM for RS.  But, we treat the $\mathbf{GP}$ as the
equivalent channel from BS to RS to design the ZF filter. According
to the principles of ZF filter~\cite{2005-DavidTse-book}, the
receiving BM based on ZF can be written as
\begin{equation}
\mathbf{F}_R=(\mathbf{P}^{\dag}\mathbf{G}^{\dag}\mathbf{G}\mathbf{P})^{-1}\mathbf{P}^{\dag}\mathbf{G}^{\dag}.
 \end{equation}
 Therefore, the linear BM at RS based on ZF-RZF can be expressed as
\begin{equation}\label{eq:F-ZFRZF}
\mathbf{F}=\mathbf{H}^{\dag}_{2}(\mathbf{H}_{2}\mathbf{H}^{\dag}_{2}+\gamma \mathbf{I})^{-1} (\mathbf{P}^{\dag}\mathbf{G}^{\dag}\mathbf{G}\mathbf{P})^{-1}\mathbf{P}^{\dag}\mathbf{G}^{\dag}.
\end{equation}
\section{Achievable Sum Rates with Base Station  and Relay Design Structures}
In this section,  we derive a quite tight lower bounds of the
achievable sum-rate of the MIMO non-regenerative relaying broadcast
network with the proposed linear transmission strategy.

After substituting the BM $\mathbf{F}$ (\ref{eq:F-ZFRZF}) in to (\ref{eq:Relay-Downlink}), the  corresponding received
signal vector at the users can be written as (\ref{eq:Downlink-ZF}).
%
%
%
From  (\ref{eq:Downlink-P-y1}) and (\ref{eq:Downlink-ZF}),  to
evaluate the amount of desired signals and interference signals at
the $k$th user, we can write the received signals at the $k$th user
during two phases in vector form as in (\ref{eq:vector-yk}),
\begin{figure*}[ht]
\begin{IEEEeqnarray} {rCl}\label{eq:Downlink-ZF}
\mathbf{y}_{2} &=&\rho_{r}\rho_{s} \mathbf{H}_{2} \mathbf{H}^{\dag}_{2}(\mathbf{H}_{2}\mathbf{H}^{\dag}_{2}+\gamma \mathbf{I})^{-1}\mathbf{s } +
\rho_{r}\mathbf{H}_{2}\mathbf{H}^{\dag}_{2}(\mathbf{H}_{2}\mathbf{H}^{\dag}_{2}+\gamma \mathbf{I})^{-1} (\mathbf{P}^{\dag}\mathbf{G}^{\dag}\mathbf{G}\mathbf{P})^{-1}\mathbf{P}^{\dag}\mathbf{G}^{\dag}\mathbf{n}_{r}+\mathbf{n}_{2}. \end{IEEEeqnarray}
\hrulefill
\begin{IEEEeqnarray} {rCl}  \label{eq:vector-yk}
\underbrace{\left[
         \begin{array}{c}
           y_{1k} \\
           y_{2k} \\
         \end{array}
       \right]}_{\mathbf{y}[k]}
&=&\underbrace{\left[
     \begin{array}{c}
       \rho_{s}\mathbf{h}_{1k}\mathbf{p}_{k} \\
       \rho_{r}\rho_{s}\mathbf{h}_{2k}\mathbf{f}_{Tk} \\
     \end{array}
   \right]}_{\mathbf{A}}s_{k}+\underbrace{\left[
                  \begin{array}{ccc}
                   \rho_{s}\mathbf{h}_{1j}\mathbf{p}_{j}&\ldots&\rho_{s}\mathbf{h}_{1K}\mathbf{p}_{K}  \\
                  \rho_{r}\rho_{s}\mathbf{h}_{2j}\mathbf{f}_{Tj}&\ldots&\rho_{r}\rho_{s}\mathbf{h}_{2K}\mathbf{f}_{TK}\\
                  \end{array}
                \right]}_{\mathbf{B}}\left[
         \begin{array}{c}
           s_{j\neq k} \\
           \vdots\\
         \end{array}
       \right]+\underbrace{\left[
                  \begin{array}{c}
                   n_{1k} \\
                   \rho_{r}\mathbf{h}_{2k}\mathbf{F}\mathbf{n}_{r}+n_{2k} \\
                  \end{array}
                \right]}_{\mathbf{N}}.
\end{IEEEeqnarray}
\hrulefill
\end{figure*}
where the second additive term $\mathbf{B}$ is the interference
signals and the third additive  term $\mathbf{N}$ is the noise
signals. Since the channel is memoryless, the average mutual
information of the $k$th user satisfies~\cite{2004-Laneman}
\begin{equation}\label{eq:MI-ZF-k}
\mathbf{I}(s_{k};\mathbf{y}[k])\leq\frac{1}{2}\log\det\left( \mathbf{I}_{2}+\mathbf{AA}^{\dag}(\mathbf{BB}^{\dag}+\textmd{E}[\mathbf{NN}^{\dag}])^{-1} \right),
\end{equation}
with equality for $s_{k}$ satisfying zero-mean, circularly symmetric complex Gaussian.
Note that
\begin{IEEEeqnarray}{rCl}\label{eq:AA}
\mathbf{AA}^{\dag}&=& \left[
               \begin{array}{cc}
                 \rho^2_{s}\mathbf{h}_{1k}\mathbf{p}_{k}\mathbf{p}^{\dag}_{k}\mathbf{h}^{\dag}_{1k} & \rho_{r}\rho^2_{s}\mathbf{h}_{1k}\mathbf{p}_{k}\mathbf{f}^{\dag}_{Tk}\mathbf{h}^{\dag}_{2k} \\
                \rho_{r}\rho^2_{s}\mathbf{h}_{2k}\mathbf{f}_{Tk}\mathbf{p}^{\dag}_{k}\mathbf{h}^{\dag}_{1k} & \rho^2_{r}\rho^2_{s}\mathbf{h}_{2k}\mathbf{f}_{Tk}\mathbf{f}^{\dag}_{Tk}\mathbf{h}^{\dag}_{2k} \\
               \end{array}
             \right],
  \nonumber\\
\mathbf{BB}^{\dag}&=&\left[
               \begin{array}{cc}
                \sum\rho^2_{s} \mathbf{h}_{1j}\mathbf{p}_{j}\mathbf{p}^{\dag}_{j}\mathbf{h}^{\dag}_{1j} &
                 \sum\rho_{r}\rho^2_{s} \mathbf{h}_{1j}\mathbf{p}_{j}\mathbf{f}^{\dag}_{Tj}\mathbf{h}^{\dag}_{2j}   \\
                 \sum\rho_{r}\rho^2_{s}           \mathbf{h}_{2j}\mathbf{f}_{Tj}\mathbf{p}^{\dag}_{j}\mathbf{h}^{\dag}_{1j}&
                  \sum\rho^2_{r}\rho^2_{s} \mathbf{h}_{2j}\mathbf{f}_{Tj}\mathbf{f}^{\dag}_{Tj}\mathbf{h}^{\dag}_{2j}\\
               \end{array}
             \right],\nonumber\\
\textmd{E}[\mathbf{NN}^{\dag}]&=&\left[
               \begin{array}{cc}
                \sigma^2_{n} &
                 0  \\
                 0&
                 \sigma^2_{n}\rho^2_{r}(\mathbf{h}_{2k}\mathbf{FF}^{\dag}\mathbf{h}^{\dag}_{2k})+\sigma^2_{n} \\
               \end{array}
             \right],\nonumber
\end{IEEEeqnarray}
where $\sum=\sum^{K}_{j=1,j\neq k}$.
It is complicated to express the expanded form of  equation~(\ref{eq:MI-ZF-k}). Here, we give  the expression of  a lower bound of the equation~(\ref{eq:MI-ZF-k}) as
\begin{IEEEeqnarray}{rCl}
\mathbf{I}(s_{k};\mathbf{y}[k])&=&\frac{1}{2}\log\det\left( \mathbf{I}_{2}+\mathbf{AA}^{\dag}(\mathbf{BB}^{\dag}+\textmd{E}[\mathbf{NN}^{\dag}])^{-1} \right)\nonumber\\
&\geq &\frac{1}{2}\log(1+\mathrm{SINR}_{1k}+\mathrm{SINR}_{2k}),\label{eq:lower}
\end{IEEEeqnarray}
where
$
\mathrm{SINR}_{1k}={\rho^2_{s}\mathbf{h}_{1k}\mathbf{p}_{k}\mathbf{p}^{\dag}_{k}\mathbf{h}^{\dag}_{1k}}/{(\sum
\mathbf{h}_{1j}\mathbf{p}_{j}\mathbf{p}^{\dag}_{j}\mathbf{h}^{\dag}_{1j}+\sigma^{2}_{n})}$,
\begin{IEEEeqnarray}{rCl}
\mathrm{SINR}_{2k}&=&\frac{\rho^2_{r}\rho^2_{s}\mathbf{h}_{2k}\mathbf{f}_{Tk}\mathbf{f}^{\dag}_{Tk}\mathbf{h}^{\dag}_{2k}}{\sum
\rho^2_{r}\rho^2_{s}
\mathbf{h}_{2j}\mathbf{f}_{Tj}\mathbf{f}^{\dag}_{Tj}\mathbf{h}^{\dag}_{2j}+\sigma^2_{n}\rho^2_{r}(\mathbf{h}_{2k}\mathbf{FF}^{\dag}\mathbf{h}^{\dag}_{2k})+\sigma^2_{n}}\nonumber
\end{IEEEeqnarray} are   the useful signal to interference plus
noise ratio (SINR) for the $k$th user during the first and second
phases, respectively.  The last step (\ref{eq:lower}) is obtained  by considering that the interference signal received by user $k$ during the first phase is irrelevant to  the interference signal  during the second phase, i.e., $\mathbf{BB}^{\dag}=\mathrm{diag}(\sum \mathbf{h}_{1j}\mathbf{p}_{j}\mathbf{p}^{\dag}_{j}\mathbf{h}^{\dag}_{1j},\sum \mathbf{h}_{2j}\mathbf{f}_{Tj}\mathbf{f}^{\dag}_{Tj}\mathbf{h}^{\dag}_{2j})$,  which will lead to interference strengthen.
Thus, the sum-rate of the network can be
expressed as
\begin{equation}\label{eq:sum-rate}
\mathbf{I}(\mathbf{s},\mathbf{y})=\sum^{K}_{k=1}\mathbf{I}(s_{k};\mathbf{y}[k]).
\end{equation}
Form the next section,  we can see that the lower bound of the
network with the proposed strategy is quite tight.

\section{Simulation Results}
In this section, the performance of the the proposed linear
transmission strategy  with an RZF at BS and a ZF combining an RZF
at RS~(RZF+ZF\& RZF) will be evaluated by using Monte Carlo
simulation with 2000 random  channel realizations.  We compare the
proposed strategy with  several different schemes without
considering DC in design, in terms of  the ergodic sum-rate of the
MIMO relaying broadcast network. Noting that, the sum-rate of  other schemes
also include the DC contribution for fair comparison. These
alternative schemes are:
\begin{enumerate}
  \item SVD-MF scheme~\cite{2010-WeiXu}:  assuming that $\mathbf{G}=\mathbf{U}\mathbf{\Lambda} \mathbf{V}^{\dag}$, then $\mathbf{P}=\sqrt{\frac{P_s}{M_b}}\mathbf{V}$, $\mathbf{F}=\rho_{r} \mathbf{F}_{T}\mathbf{U}^{\dag}$, where $\mathbf{F}_{T}=[\mathbf{h}^{\dag}_{21}/\|\mathbf{h}_{21}\|~\ldots \mathbf{h}^{\dag}_{2K}/\|\mathbf{h}_{2K}\|]$ and $\rho_{r}$ is a control factor to guarantee the power constraint in (\ref{eq:rho-r})
  \item SVD-ZF scheme~\cite{2010-WeiXu}: $\mathbf{F}_{T}=\mathbf{H}^{\dag}_{2}(\mathbf{H}_{2}\mathbf{H}^{\dag}_{2})^{-1}$, other conditions  are  the same as SVD-MF scheme.
  \item SVD-RZF scheme~\cite{2011-WZJ}: $\mathbf{F}_{T}=\mathbf{H}^{\dag}_{2}(\mathbf{H}_{2}\mathbf{H}^{\dag}_{2}+\gamma \mathbf{I})^{-1}$, where $\gamma=K\sigma^{2}_{n}/P_r$. Other conditions  are the same as SVD-MF scheme.
  \item I-MMSE scheme~\cite{2010-BinZhang}: $\mathbf{P}=\rho_{s}\mathbf{I}_{M_b}$, and $\mathbf{F}=\rho_{r}(\mathbf{H}^{\dag}_{2}\mathbf{H}_{2}+\frac{\sigma^2_{n}}{\mathbf{P}_{r}}
      \mathbf{I}_{M_r})^{-1}\mathbf{H}^{\dag}_{2}\mathbf{H}^{\dag}_{1}(\mathbf{H}^{\dag}_{1}\mathbf{H}_{1}+\sigma^2_{n}\mathbf{I}_{M_R})^{-1}$. Parameters $\rho_{s}$ and $\rho_{r}$ are scalar factors to make the power constraints at BS and RS, respectively.
  \end{enumerate}

\begin{figure}[!t]
\begin{center}
\includegraphics [width=3.2in]{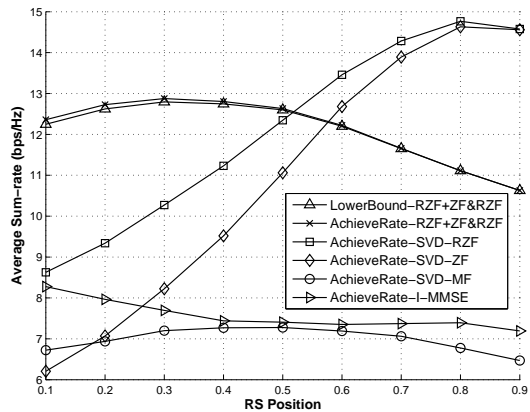}
\caption{ Average sum-rate versus  RS position~(between 0 and 1.0 ),
where BS is at $0$ point and users are at $1.0$ point,
$M_b=M_r=K=4$, and $P_s=P_r=28\mathrm{dB}$ }
\label{Fig-SumRateVsDistance}
\end{center}
\end{figure}

\begin{figure}[!t]
\begin{center}
\includegraphics [width=3.2in]{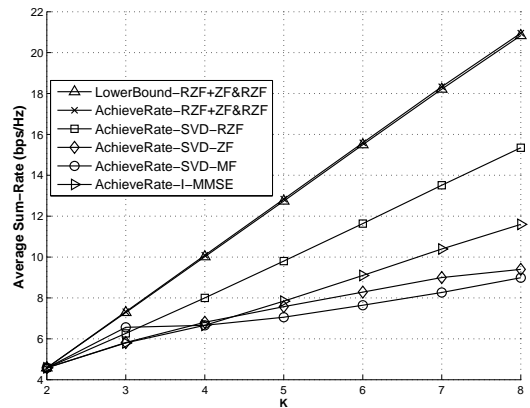}
\caption{Average sum-rate versus $K$, where $K=M_b=M_r$,
$P_s=P_r=28\mathrm{dB}$, and BS is at 0 point, RS is at 0.25 point,
users are at 1.0 point.}  \label{Fig-SumRateVsAntanna}
\end{center}
\end{figure}

All schemes are compared under the same condition of  various
network parameters and do not consider the optimal power allocation
at BS and RS. In these simulations, we consider that BS and RS are
deployed in a line with users, where all the users are deployed at
the same point. The channel gains are modeled as the combination of
large scale fading~(related to distance) and  small scale
fading~(Rayleigh fading), and all channel matrices have
\emph{i.i.d.} $\mathcal{CN}(0,\frac{1}{L^{\tau}})$ entries, where
$L$ is the distance between two nodes, and $\tau=3$ is the path loss
exponent.
\begin{figure}[!t]
\begin{center}
\includegraphics [width=3.2in]{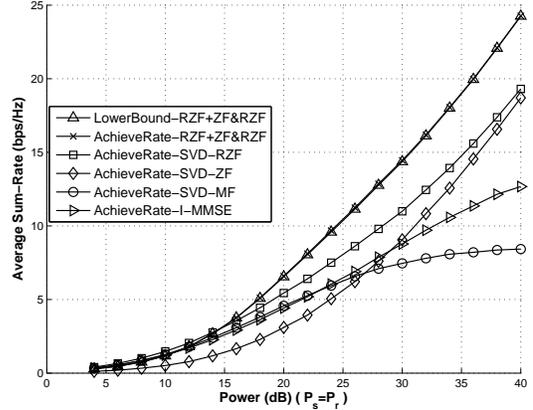}
\caption{Average sum-rate versus the transmitting power with
$P_s=P_r$, where  $M_b=M_r=K=4$, and BS is at 0 point, RS is at 0.25
point, users are at 1.0 point  } \label{Fig-SumRateVsPower}
\end{center}
\end{figure}
Fig.~\ref{Fig-SumRateVsPower}  shows the average sum-rate of the
network versus the transmitting power, when  all nodes positions are
fixed. It is observed that the sum-rate offered  by the proposed
linear transmission strategy are higher than  those by the other
linear schemes at all SNR regime, when the RS is at position of a
quarter of the distance from BS to users.
Fig.~\ref{Fig-SumRateVsDistance}  shows the average sum-rate of the
network versus the RS's position, when the powers at BS and RS are
fixed. We can see that the  average sum-rate of the proposed
strategy is higher than those of the other linear schemes, when the
RS is close to BS.
From Fig.~\ref{Fig-SumRateVsAntanna},  the average sum rate gap
between the proposed strategy and other linear schemes become larger
when the number of antennas at BS  and the number of users both
increase simultaneously. From
Fig.~\ref{Fig-SumRateVsPower}-\ref{Fig-SumRateVsAntanna}, it is
clear that the proposed strategy outperforms other linear schemes
without considering the DC in design, when the RS is close to BS.
This is because that the BM  schemes without considering DC  in
design cause $\mathrm{SINR}_{1k}$ loss (near to 0), which  leads to
rate loss when the channel gains of the DC and FC are considerable
at the scenario that  RS is close to BS. However, both SVD-RZF and
SVD-ZF schemes  outperform the proposed strategy when the RS is
close to users, because: 1) the BC is bottleneck when the RS is
close to users, and the FC is bottleneck when the RS is close to BS,
vice versa, 2) the gain of BC by SVD surpasses the loss of the DC by
SVD when the BC is bottleneck, and 3)~the ZF receiving filter at RS
will amplify the noise, especially at the case that the RS is close
to users, which leads to rate loss of the proposed strategy.
\section{Conclusion and discussions}
In this letter,  we propose a linear transmission strategy to design
PM at BS and BM at RS for the MIMO relaying broadcast network with
DC.  In the proposed scheme, we take an RZF filter based on DC as
the PM at BS, an RZF filter based on FC as transmitting BM at RS,
and  a ZF filter as  receiving BM at RS based on the PM at BS and
BC. Numerical results shown that  the proposed strategy outperforms
the other linear schemes without considering DC in design, when the
RS is  close to BS.
In fact, the proposed linear strategy is a general scheme to design
the PM at BS and BM at RS with DC. One can choose other linear
precoders at BS and RS instead of the RZF, such as the precoder
based on SLNR~\cite{2007-Mirette}, and other relay linear receiving
filter can be chosen as receiving BM at RS, such as
MMSE~\cite{2005-DavidTse-book} filter and so on.





\bibliography{mybib}
\bibliographystyle{ieeetr}
%


%

\end{document}